\documentclass[fleqn,10pt]{article}
\usepackage{epsfig}
\newcommand{\be}{\begin{equation}}
\newcommand{\ee}{\end{equation}}
\newcommand{\ba}{\begin{eqnarray}}
\newcommand{\ea}{\end{eqnarray}}

\newcommand{\eq}[1]{Eq.\,(\ref{#1})}

\newcommand{\delchi}{\Delta \chi^2_i}

\newcommand{\delchimax}{{\delchi}_{\rm max}}
\newcommand{\x}{(\nu/m)}
\newcommand{\y}{(\nu_0/m)}
\begin{document}
\setcounter{secnumdepth}{4}
\renewcommand\thepage{\ }
%
%
\begin{titlepage} 
%
\newcommand\reportnumber{1261} 
\newcommand\mydate{\today} 
\newlength{\nulogo} 
\settowidth{\nulogo}{\small\sf{NUHEP Report XXX}}
\title{
\vspace{-.8in} 
\hfill\fbox{{\parbox{\nulogo}{\small\sf{
NUHEP Report  \reportnumber\\
Brown-HET-1465
          \mydate}}}}
\vspace{0.5in} \\
{
New limits on ``odderon'' amplitudes from analyticity constraints
}}

\author{
Martin~M.~Block\\
{\small\em Department of Physics and Astronomy,} \vspace{-5pt} \\ 
{\small\em Northwestern University, Evanston, IL 60208}\\
\vspace{-5pt}
\  \\
Kyungsik~Kang
\vspace{-5pt} \\ 
{\small\em Department of Physics,} 
\vspace{-5pt} \\ 
{\small\em Brown University, Providence,  RI, 02912} \\
\vspace{-5pt}\\
%
\vspace{-5pt}\\
%
}    
\vspace{.5in}
\vfill
\date {}
\maketitle
\begin{abstract}
In studies of high energy $pp$ and $\bar pp$ scattering, the odd (under crossing) forward scattering amplitude accounts for the difference between the $pp$ and $\bar pp$ cross sections.  Typically, it is taken as $f_-=-\frac{p}{4\pi}Ds^{\alpha-1}e^{i\pi(1-\alpha)/2}$ ($\alpha\sim 0.5$), which has $\Delta\sigma, \Delta\rho\rightarrow0$ as $s\rightarrow\infty$, where $\rho$ is the ratio of the real to the imaginary portion of the forward scattering amplitude. However, the odd-signatured amplitude can have in principle a strikingly different behavior, ranging from having $\Delta\sigma\rightarrow$non-zero constant to having $\Delta\sigma \rightarrow \ln s/s_0$ as $s\rightarrow\infty$, the maximal behavior allowed by analyticity and the Froissart bound. We reanalyze high energy $pp$ and $\bar pp$ scattering data, using new analyticity constraints, in order to put new and precise limits on the magnitude of ``odderon'' amplitudes.

\end{abstract}
\end{titlepage} 
%
\pagenumbering{arabic}
\renewcommand{\thepage}{-- \arabic{page}\ --}  
The conventional odd (under crossing) laboratory forward scattering amplitude  used for $pp$ and $\bar pp$ scattering, suggested by Regge theory, is
\be
\frac{4\pi}{p}f_-=-Ds^{\alpha-1}e^{i\pi(1-\alpha)/2},
\ee 
which results in $\Delta\sigma\equiv\sigma_{pp}-\sigma_{\bar pp}\rightarrow 0, \Delta\rho\equiv\rho_{pp}-\rho_{\bar pp} \rightarrow 0$ as $s\rightarrow\infty$. Nicolescu et al\cite{nicolescu1,nicolescu2,joynson} have introduced odd amplitudes called ``odderons'',  with the interesting properties that they can have $\Delta\sigma\rightarrow$non-zero constant to even having $\Delta\sigma \rightarrow \ln s/s_0$ as $s\rightarrow\infty$. 

There has been mounting evidence from many sources that the crossing-even hadron-hadron cross section behaves at high energy as $\ln^2 s$, thus saturating the Froissart bound, a result with a rather profound physical significance. Using factorization and {\em simultaneously} fitting real analytic forward scattering amplitudes to $\gamma\gamma$ cross sections, $\gamma p$ cross sections and $pp$ and $\bar pp$ cross sections and $\rho$-values, Block and Kang\cite{bk} have shown  that a $\ln^2 s$ fit, saturating the Froissart bound, is in accord with the experimental data. The COMPETE group\cite{cudell}, globally fitting hadron-hadron cross sections, has offered evidence that favors a $\ln^2 s$ behavior at high energies. Igi and Ishida\cite{igiandishidapip,igiandishidapp} have shown that the $\pi^\pm p$ systems and the $pp$ and $\bar pp$ systems  saturate the Froissart bound, using finite energy sum rules. Kang and Nastase\cite{blackhole} proved that saturation of the QCD Froissart bound is related to the creation of black holes of AdS size in Planckian scattering. Block and Halzen have shown that the Froissart bound is saturated for the $\gamma p$ system\cite{BH}, the $\pi^\pm p$ systems and the $pp$ and $\bar pp$ systems\cite{bhpp}, i.e.,  the even (under crossing) cross section rose asymptotically as $\ln^2s$. For their nucleon-nucleon analysis they used 4 analyticity constraints that anchored the high energy cross section parametrizations to both the experimental $pp$ and $\bar pp$ cross sections and their first derivatives at $\sqrt s=4$ GeV, giving fits with the smallest statistical parameter errors. This technique completely ruled out the possibility of an asymptotic $\ln s$ rise.  In this communication we  extend their analysis to include ``odderons''.  

Block and Cahn\cite{bc} made an odderon analysis of $pp$ and $\bar pp$ scattering in 1985 that put limits on odderon amplitudes. Since we will later want to directly compare our results with theirs, we will use their notation.  Using forward real analytic amplitudes to describe the data, they wrote\cite{bc} the  crossing-even real analytic laboratory amplitude for forward high energy scattering as
\begin{equation}
\frac{4\pi}{p}f_+=i\left\{A+\beta[\ln (s/s_0) -i\pi/2]^2+cs^{\mu-1}e^{i\pi(1-\mu)/2}-i\frac{4\pi}{p}f_+(0)\right\},\label{evenamplitude_gp}
\end{equation}
and the conventional crossing-odd real analytic forward amplitude as 
\be
\frac{4\pi}{p}f_-=-Ds^{\alpha -1}e^{i\pi(1-\alpha)/2}.\label{oddamplitude}
\ee
Here $\alpha < 1$ parametrizes the Regge behavior of the crossing-odd amplitude which vanishes at high energies and $A$, $\alpha$, $\beta$, $c$, $D$, $s_0$ and $\mu$ are real constants. The variable $s$ is the square of the center of mass system (c.m.) energy,  $p$ is the laboratory momentum.
 The additional real constant $f_+(0)$ is the subtraction constant at $\nu=0$ needed to be introduced in a singly-subtracted dispersion relation\cite{bc},\cite{gilman}.  

Again, following Block and Cahn\cite{bc}, we now introduce three types of odderon laboratory amplitudes for forward scattering, $f_-^{(j)}$, where $j=0,1,$ or 2. Introducing the laboratory energy $\nu=\sqrt{p^2+m^2}$, where $m$ is the proton mass, they are:
\ba
 f_-^{(0)}&=&-\frac{1}{4\pi}\epsilon^{(0)}\nu\label{odd0},\\
 f_-^{(1)}&=&-\frac{1}{4\pi}\epsilon^{(1)}\nu\left[\ln (s/s_0)-i\frac{\pi}{2}\right]\label{odd1},\\
 f_-^{(2)}&=&-\frac{1}{4\pi}\epsilon^{(2)}\nu\left[\ln (s/s_0)-i\frac{\pi}{2}\right]^2\label{odd2},
\ea 
where the $\epsilon^{(j)}\ j=0,1,2$ are all real coefficients. These amplitudes, called odderon 0, odderon 1 and odderon 2, respectively, are manifestly odd, since they are all  proportional to $\nu$ times an even amplitude. Clearly, the laboratory energy $\nu$  is odd under crossing ($\nu\rightarrow -\nu$), whereas terms like $[\ln(s/s_0)-i\frac{\pi}{2}]$  are even under crossing, so that their overall product, $f_-^{(j)}$,  is crossing-odd.  It can be shown that odderon 2 is the `maximal' odderon allowed by unitarity and the Froissart bound (see Eqns. (4.114) and (4.115) of Ref. \cite{bc}). We will combine these odderons individually with the conventional odd amplitude of \eq{oddamplitude} to form a new total odd amplitude.  Since it is pure real, the amplitude $f_-^{(0)}$  only causes a small splitting in the $\rho$-values at high energy; the amplitude $f_-^{(1)}$ has a constant imaginary part, so that it leads to a constant non-zero $\Delta\sigma$, while its real part causes the $\rho$-values to split apart at high energy ; finally, the amplitude $f_-^{(2)}$  has an imaginary part  that  causes $\Delta\sigma\rightarrow\ln(s/s_0)$ as $s \rightarrow\infty$, along with a real part that causes a substantial splitting of the $\rho$-values at high energies. We have chosen these amplitudes to be identical to those that were used by Block and Cahn\cite{bc} in their work,  so that at the  end of our analysis we can make a direct comparison of our odderon coefficients $\epsilon ^{(j)}$ with theirs.  We comment that that these real analytic forward scattering amplitudes, \eq{evenamplitude_gp}--\eq{odd2}, can also be derived as solutions to derivative dispersion relations\cite{nicolescu2}.

Using the optical theorem and our laboratory forward scattering amplitude normalization, we write
\ba
\sigma_{\rm even}&=&\frac{4\pi}{p}{\rm Im}\,f_+\\
\sigma_{\rm odd}&=&\frac{4\pi}{p}{\rm Im}\,f_-\label{sigma+-}, 
\ea
the even and odd (under crossing) cross sections due to the even and odd forward laboratory amplitudes $f_+$ and $f_-$, respectively. These cross section sums and differences
\ba
\sigma(pp)&\equiv&\sigma_{\rm even}+\sigma_{\rm odd},\\
\sigma(\bar pp)&\equiv&\sigma_{\rm even}-\sigma_{\rm odd},
\ea 
give rise to the $pp$ and the $\bar pp$ cross sections, respectively.

We remind the reader that the optical theorem states that the cross section contributions of the amplitudes of \eq{odd0}, \eq{odd1} and \eq{odd2} are obtained by multiplying ${\rm Im}\,f_-^{(j)}$ by $4\pi/p$. Thus, we see that what is needed to  combine an odderon amplitude with the normal amplitude is the term  $\frac{4\pi}{p}f_-^{(j)}$. Using the optical theorem and analyticity in the high energy limit where $p=\nu$---after noting that $\frac{4\pi}{p}f_-^{(j)}$ can be replaced by $\frac{4\pi}{\nu}f_-^{(j)}$---we obtain the total cross sections $\sigma^\pm_{(j)}$ and $\rho^\pm_{(j)}$, the ratios of the real to the imaginary portion of the forward scattering amplitude, for  $j=0,1,2$, as
\ba
\sigma^{\pm}_{(0)}&=& A+\beta\left[\ln^2 s/s_0-\frac{\pi^2}{4}\right]+c\,\sin(\pi\mu/2)s^{\mu-1}\pm D\cos(\pi\alpha/2)s^{\alpha -1},  \label{sigeps0}\\
\rho^\pm_{(0)}&=&{1\over\sigma^\pm_{(0)}}\left\{\beta\,\pi\ln s/s_0-c\,\cos(\pi\mu/2)s^{\mu-1}+\frac{4\pi}{\nu} f_+(0)\pm D\sin(\pi\alpha/2)s^{\alpha -1}\pm\epsilon^{(0)}\right\},\label{rhoeps0}
\ea 
or
\ba
\sigma^{\pm}_{(1)}&=& A+\beta\left[\ln^2 s/s_0-\frac{\pi^2}{4}\right]+c\,\sin(\pi\mu/2)s^{\mu-1}\pm D\cos(\pi\alpha/2)s^{\alpha -1}\mp\epsilon^{(1)}\frac{\pi}{2},  \label{sigeps1}\\
\rho^\pm_{(1)}&=&{1\over\sigma^\pm_{(1)}}\left\{\beta\,\pi\ln s/s_0-c\,\cos(\pi\mu/2)s^{\mu-1}+\frac{4\pi}{\nu} f_+(0)\pm D\sin(\pi\alpha/2)s^{\alpha -1}\pm\epsilon^{(1)}\ln(s/s_0)\right\},\label{rhoeps1}
\ea
or
\ba
\sigma^{\pm}_{(2)}&=& A+\beta\left[\ln^2 s/s_0-\frac{\pi^2}{4}\right]+c\,\sin(\pi\mu/2)s^{\mu-1}\pm D\cos(\pi\alpha/2)s^{\alpha -1}\mp\epsilon^{(2)}\pi\ln(s/s_0),  \label{sigeps2}\\
\rho^\pm_{(2)}&=&{1\over\sigma_\pm^{(2)}}\left\{\beta\,\pi\ln s/s_0-c\,\cos(\pi\mu/2)s^{\mu-1}+\frac{4\pi}{\nu} f_+(0)\pm D\sin(\pi\alpha/2)s^{\alpha -1}\right.\\
&&\left.\qquad\qquad\qquad\qquad\qquad\qquad\qquad\qquad\qquad\qquad\qquad\qquad\pm\epsilon^{(2)}\left(\ln^2(s/s_0)-\frac{\pi^2}{4}\right)\right\},\label{rhoeps2}
\ea
where the  upper sign is for $pp$ and the lower sign is for $\bar p p$ scattering.  

We now introduce the definitions \ba A& =& c_0 + \frac{\pi^2}{4}c_2 - \frac{c_1 ^ 2}{ 4c_2}, \\
s_0 &= &2m ^ 2 e^{-c_1 / (2c_2)},\label{s0}\\
\beta&=&c_2,\\
c &=& \frac{(2m^2)^{1 - \mu} } {\sin(\pi\mu/ 2)}\beta_{\cal P'},\\ 
D&=&\frac{(2m^2)^{1-\alpha}}{\cos(\pi\alpha/2)}\delta.
\ea
After some algebraic manipulations, the cross sections $\sigma^\pm_{(j)}$ and the $\rho$-values  $\rho^\pm_{(j)}$, along with the cross section derivatives $\frac{d\sigma^{\pm}_{(j)}}{d\x}$,  can now be written as
\begin{eqnarray}
\sigma^\pm_{(0)}(\nu)&{\!\!\! =\!\!\! }&c_0+c_1\ln\left(\frac{\nu}{m}\right)+c_2\ln^2\left(\frac{\nu}{m}\right)+\beta_{\cal P'}\left(\frac{\nu}{m}\right)^{\mu -1}\pm\  \delta\left({\nu\over m}\right)^{\alpha -1},\label{sigmapm0}\\
\rho^\pm_{(0)}(\nu)&{\!\!\! =\!\!\! }&{1\over\sigma^\pm_{(0)}}\left\{\frac{\pi}{2}c_1+c_2\pi \ln\left(\frac{\nu}{m}\right)-\beta_{\cal P'}\cot\left({\pi\mu\over 2}\right)\left(\frac{\nu}{m}\right)^{\mu -1}+\frac{4\pi}{\nu}f_+(0)\right.\nonumber\\
&&\qquad\qquad\qquad\qquad\qquad\qquad\qquad\qquad\left.\pm \delta\tan\left({\pi\alpha\over 2}\right)\left({\nu\over m}\right)^{\alpha -1} \pm\epsilon^{(0))}\right\}\!\!,\label{rhopm0}\\
\frac{d\sigma^{\pm}_{(0)}(\nu)}{d\x}&{\!\!\! =\!\!\! }&c_1\left\{\frac{1}{\x}\right\} +c_2\left\{ \frac{2\ln\x}{\x}\right\}+\beta_{\cal P'}\left\{(\mu-1)\x^{\mu-2}\right\}\label{derivpm0} 
\pm \ \delta\left\{(\alpha -1)\x^{\alpha - 2}\right\}
\end{eqnarray}
or
\begin{eqnarray}
\sigma^\pm_{(1)}(\nu)&{\!\!\! =\!\!\! }&c_0+c_1\ln\left(\frac{\nu}{m}\right)+c_2\ln^2\left(\frac{\nu}{m}\right)+\beta_{\cal P'}\left(\frac{\nu}{m}\right)^{\mu -1}\pm\  \delta\left({\nu\over m}\right)^{\alpha -1}\mp\epsilon^{(1)}\frac{\pi}{2},\label{sigmapm1}\\
\rho^\pm_{(1)}(\nu)&{\!\!\! =\!\!\! }&{1\over\sigma^\pm_{(1)}}\left\{\frac{\pi}{2}c_1+c_2\pi \ln\left(\frac{\nu}{m}\right)-\beta_{\cal P'}\cot\left({\pi\mu\over 2}\right)\left(\frac{\nu}{m}\right)^{\mu -1}+\frac{4\pi}{\nu}f_+(0)\right.\nonumber\\
&&\qquad\qquad\qquad\qquad\qquad\qquad\qquad\qquad\left.\pm \delta\tan\left({\pi\alpha\over 2}\right)\left({\nu\over m}\right)^{\alpha -1} \pm\epsilon^{(1)}\ln(s/s_0)\right\}\!\!,\label{rhopm1}\\
\frac{d\sigma^{\pm}_{(1)}(\nu)}{d\x}&{\!\!\! =\!\!\! }&c_1\left\{\frac{1}{\x}\right\} +c_2\left\{ \frac{2\ln\x}{\x}\right\}+\beta_{\cal P'}\left\{(\mu-1)\x^{\mu-2}\right\}\label{derivpm1} 
\pm \ \delta\left\{(\alpha -1)\x^{\alpha - 2}\right\}
\end{eqnarray}
or
\begin{eqnarray}
\sigma^\pm_{(2)}(\nu)&{\!\!\! =\!\!\! }&c_0+c_1\ln\left(\frac{\nu}{m}\right)+c_2\ln^2\left(\frac{\nu}{m}\right)+\beta_{\cal P'}\left(\frac{\nu}{m}\right)^{\mu -1}\pm\  \delta\left({\nu\over m}\right)^{\alpha -1}\mp\epsilon^{(2)}\pi\ln(s/s_0),\label{sigmapm2}\\
\rho^\pm_{(2)}(\nu)&{\!\!\! =\!\!\! }&{1\over\sigma^\pm_{(2)}}\left\{\frac{\pi}{2}c_1+c_2\pi \ln\left(\frac{\nu}{m}\right)-\beta_{\cal P'}\cot\left({\pi\mu\over 2}\right)\left(\frac{\nu}{m}\right)^{\mu -1}+\frac{4\pi}{\nu}f_+(0)\right.\nonumber\\
&&\qquad\qquad\qquad\qquad\left.\pm \delta\tan\left({\pi\alpha\over 2}\right)\left({\nu\over m}\right)^{\alpha -1} \pm\epsilon^{(2)}\left(\ln^2(s/s_0)-\frac{\pi^2}{4}\right)\right\}\!\!,\label{rhopm2}\\
\frac{d\sigma^{\pm}_{(2)}(\nu)}{d\x}&{\!\!\! =\!\!\! }&c_1\left\{\frac{1}{\x}\right\} +c_2\left\{ \frac{2\ln\x}{\x}\right\}+\beta_{\cal P'}\left\{(\mu-1)\x^{\mu-2}\right\}\mp\epsilon^{(2)}\left\{\frac{\pi}{\x}\right\}\label{derivpm2}  \nonumber\\
&&\ \ \ \ \ \ \ \ \ \ \ \ \ \ \ \ \ \ \ \ \  
\pm \ \delta\left\{(\alpha -1)\x^{\alpha - 2}\right\},
\end{eqnarray}
in the high energy limit where $s\rightarrow 2m\nu$, where the upper sign is for $pp$ and the lower sign is for $\bar p p$ scattering.  Units of $\sigma$ in mb, and $\nu$ and $m$ in GeV, where $m$ is the proton mass, will be used.  We will use $\mu=0.5$ , the value%
\footnote{We use the value $\mu=0.5$ in order to be able to directly compare our results, using the same data set, the same high energy parametrization and the same constraints, with an analysis\cite{bhpp} which used $\epsilon^{(j)}=0,\ j=0,1,2,$ i.e., had no odderon amplitudes in its parametrization\label{same}.} %
used by Block and Halzen\cite{bhpp},  which is appropriate for a Regge-descending trajectory.  The new even coefficients $c_0$, $c_1$, $c_2,$, $\beta_{\cal P'}$ and the odd coefficient $\delta$, along with the exponents $\mu$ and $\alpha$, are all real.
 These transformations linearize  \eq{sigmapm0}, \eq{sigmapm1} and \eq{sigmapm2} in the parameters $c_0, c_1, c_2, \beta_{\cal P'}$ and $\delta$, convenient for a $\chi^2$ fit to the experimental total cross sections and $\rho$-values. 

We will use new analyticity constraints\cite{analyticity} in the fitting of the $\bar p p$ and $pp$ data that anchor the theoretical cross sections and their derivatives of our high energy parametrization with {\em experimental} cross sections and their derivatives at a transition energy $\nu_0$ which is just above the resonance region.   
Let  $\sigma^+$  and $\sigma^-$  be the total cross sections for $pp$ and $\bar p p$ scattering. It is convenient to define 4  {\em experimental} quantities evaluated at the transition energy $\nu_0$. The transition energy $\nu_0$ is a low energy after which resonance behavior finishes. Following Block and Halzen\cite{bhpp}, we will choose $\nu_0=7.59$ GeV (corresponding to $\sqrt s_0=4$ GeV). 

We now introduce 4 new well-determined experimental quantities, 2 crossing even quantities $\sigma_{\rm av}$ and $m_{\rm av}$  and  2 crossing-odd quantities $\Delta\sigma$ and $\Delta m$, 
\begin{eqnarray}
\sigma_{\rm av}&\equiv&\frac{\sigma^{+}\y+\sigma^-\y}{2},\nonumber\\
\Delta\sigma&\equiv &\frac{\sigma^{+}\y-\sigma^-\y}{2},\nonumber\\
m_{\rm av}&\equiv&\frac{1}{2}\left(\frac{d\sigma^{+}}{d\x}+\frac{d\sigma^{-}}{d\x}\right)_{\nu =\nu_0},\nonumber\\
\Delta m&\equiv&\frac{1}{2}\left(\frac{d\sigma^{+}}{d\x}-\frac{d\sigma^{-}}{d\x}\right)_{\nu =\nu_0},
\end{eqnarray}
capitalizing on the very accurate low energy experimental $pp$ and $\bar pp$ cross section data that are available.

Using  $\sigma_{\rm av}$ and $m_{\rm av}$, we now write the 2 crossing-even analyticity constraint equations as 
\begin{eqnarray}
\beta_{\cal P'}&=&\frac{\y^{2-\mu}}{\mu -1}\left[m_{\rm av}-c_1\left\{\frac{1}{\y}\right\} -c_2\left\{\frac{2\ln\y}{\y}
\right\}\right],\label{deriveven}\\
c_0&=& \sigma_{\rm av}-c_1\ln\y-c_2\ln^2\y-\beta_{\cal P'}\y^{\mu-1},\label{intercepteven}
\end{eqnarray}
reiterating that \eq{deriveven} and \eq{intercepteven} utilize the {\em experimental} even cross section $\sigma_{\rm av}$ and its slope $m_{\rm av}$ evaluated at the transition energy $\nu_0$, where we join on to the asymptotic fit. 

The situation is a little more complicated for the crossing-odd  constraints. For odderon 0, we have
\begin{eqnarray}\qquad
\alpha&=&1+\frac{\Delta m}{\Delta \sigma}\times\frac{\nu_0}{m},\ \qquad\qquad\quad\qquad\qquad\qquad j=0,\label{derivodd0}\\
\delta&=&\Delta \sigma\times\left(\frac{\nu_0}{m}\right)^{1-\alpha}\label{interceptodd0},
\end{eqnarray}
whereas for odderon 1, we find
\begin{eqnarray}
\alpha&=&1+\frac{\Delta m}{\Delta \sigma-\epsilon^{(1)}(\frac{\pi}{2})}\,\times\frac{\nu_0}{m},\qquad\qquad\quad\qquad\qquad j=1,\label{derivodd1}\\
\delta&=&\Delta \sigma\times\left(\frac{\nu_0}{m}\right)^{1-\alpha}\label{interceptodd1},
\end{eqnarray}
and for odderon 2,
\begin{eqnarray}
\alpha&=&1+\frac{\Delta m-\epsilon^{(2)}\left\{\pi\nu_0/m\right\}}{\Delta \sigma-\epsilon^{(2)}\{\pi\ln(2m\nu_0 /s_0)\}}\times\frac{\nu_0}{m},\qquad\qquad j=2,\label{derivodd2}\\
\delta&=&\Delta \sigma\times\left(\frac{\nu_0}{m}\right)^{1-\alpha}\label{interceptodd2},
\end{eqnarray}
where $s_0=22.9$ GeV$^2$, which is the approximate value of $s_0$ found from the fit parameters of Table \ref{table:ppoddfit}, using \eq{s0}. Again,  the crossing-odd  constraints  $\Delta \sigma$ and $\Delta m$ are fixed by the {\em experimental} $pp$ and $\bar pp$ cross sections and their derivatives at the transition energy $\nu_0$.


Utilizing the rich amount of  accurate low energy data at the transition energy $\nu_0$, we have now constrained our high energy fit at $\nu_0=7.59$ GeV\cite{bhpp}.   For safety, the  data fitting is started at an energy $\nu_{\rm min}=18.25$ GeV (corresponding to $\sqrt {s_{\rm min}}=6$ GeV), appreciably higher than the transition energy (see footnote \ref{same}).  The appropriate cross sections and slopes, taken from ref. \cite{bhpp} , are summarized in Table \ref{table:transitionparameters}, along with the minimum energies used in the asymptotic fits (see footnote \ref{same}). Very local fits had been  made to the region about the energy $\nu_0$ in order to evaluate the two cross sections and their two derivatives at $\nu_0$ that were needed in the above constraint equations. We next impose the 4 constraint equations  arising from analyticity\cite{analyticity}:
\begin{itemize}
 \item For Odderon 0, the Equations (\ref{deriveven}), (\ref{intercepteven}), (\ref{derivodd0}) and (\ref{interceptodd0}), are used in  our $\chi^2$ fit to Equations (\ref{sigmapm0}) and (\ref{rhopm0}).
\item 
For Odderon 1, the Equations (\ref{deriveven}), (\ref{intercepteven}), (\ref{derivodd1}) and (\ref{interceptodd1}),  are used in  our $\chi^2$ fit to Equations (\ref{sigmapm1}) and (\ref{rhopm1}).
\item
For Odderon 2, the Equations (\ref{deriveven}), (\ref{intercepteven}), (\ref{derivodd2}) and (\ref{interceptodd2})  are used in  our $\chi^2$ fit to Equations (\ref{sigmapm2}) and (\ref{rhopm2}).
\end{itemize}

 We stress that the odd amplitude parameters $\alpha$ and $\delta$ and hence the odd amplitude itself is {\em completely determined} by the experimental values $\Delta m$ and $\Delta \sigma$ at the transition energy $\nu_0$ and the value of $\epsilon^{(j)},\  j=0,1,2$.    Further,  the even amplitude parameters $c_0$ and $\beta_{\cal P}'$ are now determined by $c_1$ and $c_2$,  along with the experimental values of $\sigma_{\rm av}$ and $m_{\rm av}$ at the transition energy $\nu_0$. In particular, we only fit the 4  parameters $c_1$, $c_2$, $f_+(0)$ and $\epsilon^{(j)},\  j=0,1,2$. Since the subtraction constant $f_+(0)$  enters only into the $\rho$-value determinations, of the original 8 free parameters that were  needed to be fit for a $\ln^2s$  energy dependence of the cross sections $\sigma^{\pm}$, only the 3  parameters  $c_1$, $c_2$ and $\epsilon^{(j)},\ j=0,1,2$ are now free,   giving  us exceedingly little freedom in this fit---it is indeed very tightly constrained, with little latitude for adjustment.  

The adaptive Sieve  algorithm\cite{sieve} that minimizes the effect that ``outliers''---points with abnormally high contributions to $\chi^2$---have on a fit when they contaminate a data sample that is otherwise Gaussianly distributed is described in Refs. \cite{bhpp} and \cite{sieve}. The sieved data set that we will use for our $\chi^2$ fit to $\sigma_{pp},\ \sigma_{\bar pp},\ \rho_{pp}$ and $\rho_{\bar p p}$ for $
\sqrt s\ge 6$ GeV is detailed in Ref. \cite{bhpp}, where Block and Halzen found that the 25 points that were screened out had a $\chi^2$ contribution of $\approx 980$, an average value of $\approx 39$, using the cut $\delchimax=6$.  For a Gaussian distribution, about 3 points with $\delchi>6$ are expected, giving a total $\chi^2$ contribution of slightly more than 18 and {\em not} 980. The effect of the ``Sieve'' algorithm in ridding the data sample of outliers is major.

Table \ref{table:ppoddfit} summarizes the results of our 3 simultaneous fits to the available accelerator data,  using the sieved data set of ref. \cite{bhpp} which was obtained after using the ``Sieve'' algorithm on the Particle Data Group\cite{pdg} compendium for $\sigma_{pp}$, $\sigma_{\bar pp}$, $\rho_{pp}$ and $\rho_{\bar pp}$, using a minimum fitting energy $\sqrt {s_{\rm min}}=6$ GeV and imposing the cut $\delchimax=6$. The fits were made using 4 constraint equations with a transition energy $\sqrt s_0=4$ GeV,  for odderons 0, 1 and 2.  Very satisfactory probabilities ($\sim  0.2$) for 183 degrees of freedom were found for all 3 odderon choices.

\begin{figure}[h,t,b] 
\begin{center}
\mbox{\epsfig{file=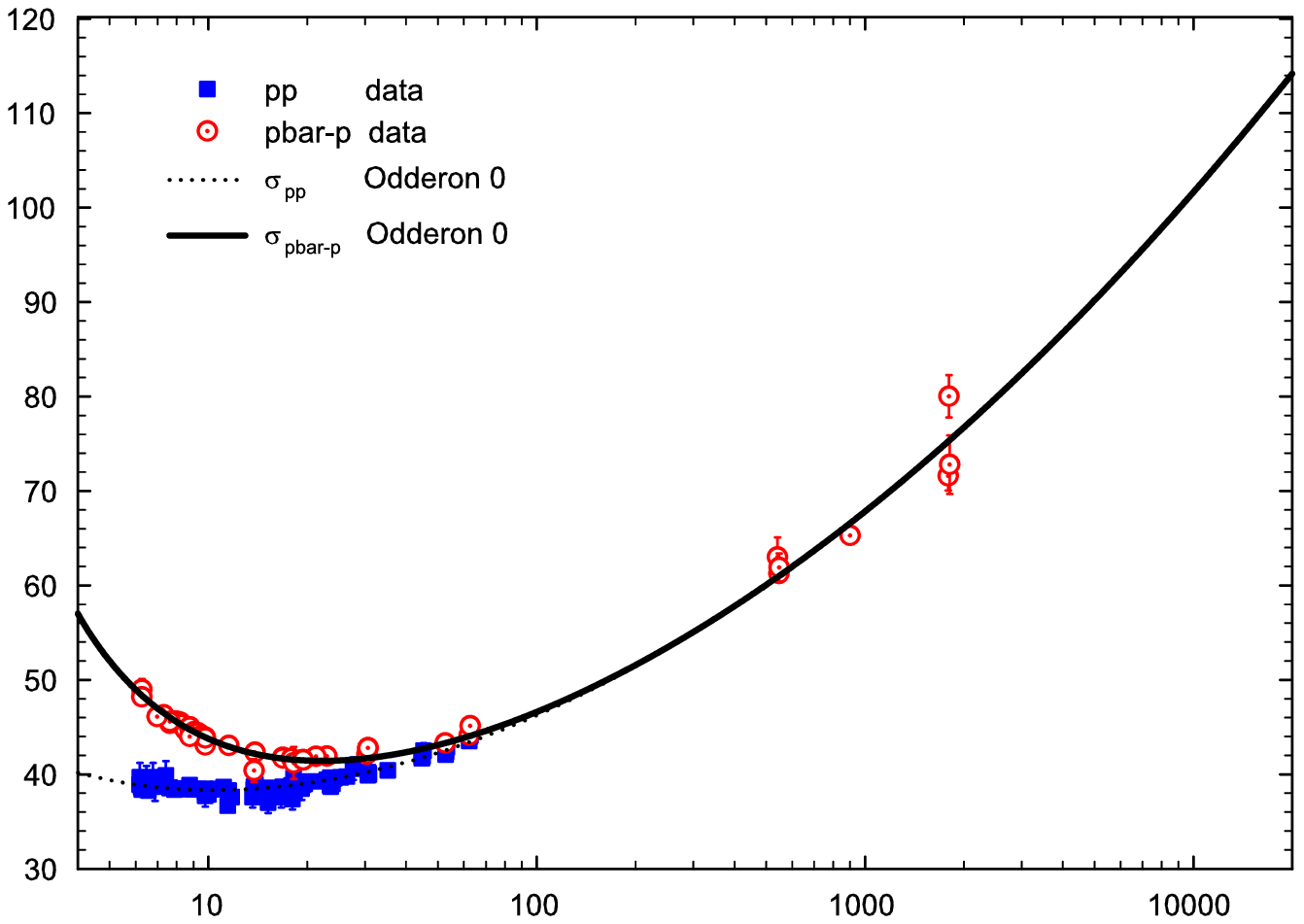,width=4.8in%
}}
\end{center}	
\caption[]{ \footnotesize Odderon 0: The fitted total cross sections $\sigma_{\bar pp}$ and $\sigma_{pp}$ in mb, vs. $\sqrt s$, in GeV, using the 4 constraints of Equations (\ref{deriveven}), (\ref{intercepteven}), (\ref{derivodd0}) and (\ref{interceptodd0}), for odderon 0 of \eq{odd0}.  The circles are the sieved  data  for $\bar pp$ scattering and the squares are the sieved data for $pp$ scattering for $\sqrt s\ge 6$ GeV. The solid curve ($\bar pp$)  and the dotted curve ($pp$) are $\chi^2$ cross section fits, corresponding to a simultaneous fit to cross sections and $\rho$-values   (Table \ref{table:ppoddfit}, of odderon 0) of  \eq{sigmapm0} and \eq{rhopm0}.
}
\label{fig:sigmaodd0}
\end{figure}

\begin{figure}[h,t,b] 
\begin{center}
\mbox{\epsfig{file=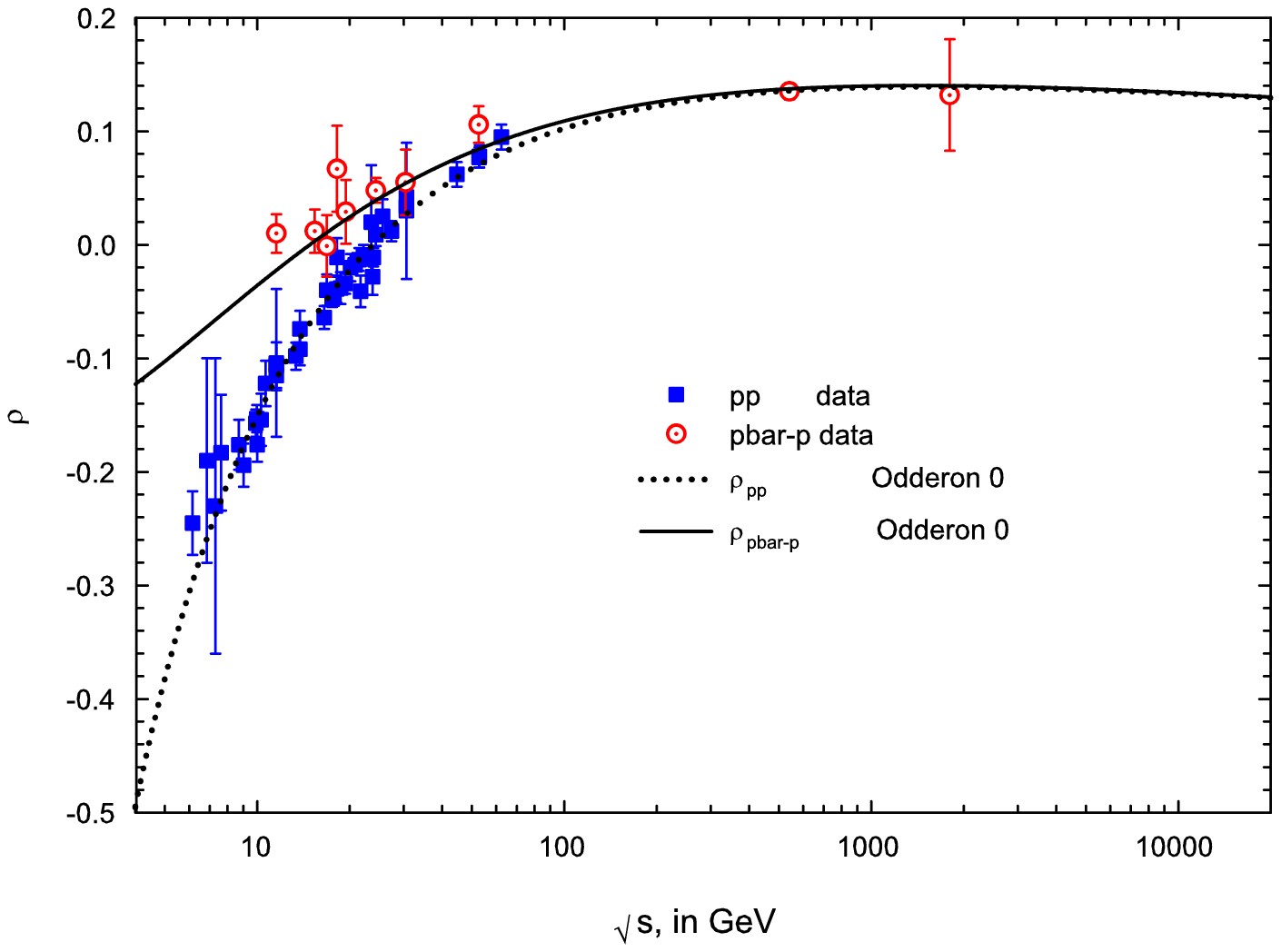
,width=4.8in,
bbllx=0pt,bblly=0pt,bburx=420pt,bbury=305pt,clip=%
}}
\end{center}
\caption[]{ \footnotesize Odderon 0: The fitted $\rho$-values, $\rho_{\bar pp}$ and $\rho_{pp}$, vs. $\sqrt s$, in GeV, using the 4 constraints of Equations (\ref{deriveven}), (\ref{intercepteven}), (\ref{derivodd0}) and (\ref{interceptodd0}), for odderon 0 of \eq{odd0}.  The circles are the sieved  data  for $\bar pp$ scattering and the squares are the sieved data for $pp$ scattering for $\sqrt s\ge 6$ GeV. The solid curve ($\bar pp$)  and the dotted curve ($pp$) are $\chi^2$ cross section fits, corresponding to a simultaneous fit to cross sections and $\rho$-values   (Table \ref{table:ppoddfit}, of odderon 0) of  \eq{sigmapm0} and \eq{rhopm0}.
}
\label{fig:rhoodd0}
\end{figure}

\begin{figure}[h,t,b] 
\begin{center}
\mbox{\epsfig{file=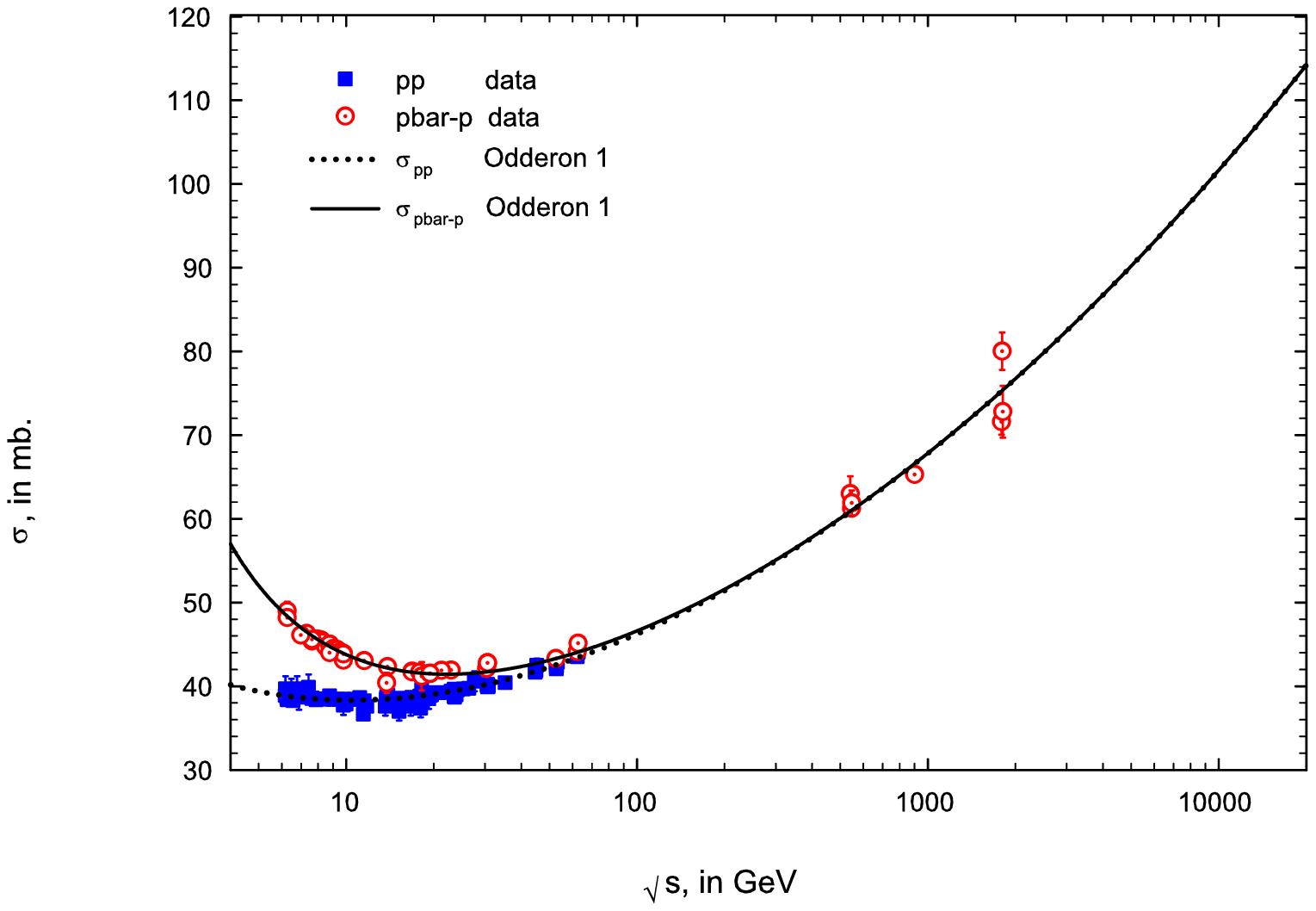,width=4.8in%
,bbllx=0pt,bblly=0pt,bburx=445pt,bbury=305pt,clip=%
}}
\end{center}
\caption[]{ \footnotesize
Odderon 1: The fitted total cross sections $\sigma_{\bar pp}$ and $\sigma_{pp}$ in mb, vs. $\sqrt s$, in GeV, using the 4 constraints of Equations (\ref{deriveven}), (\ref{intercepteven}), (\ref{derivodd1}) and (\ref{interceptodd1}), for odderon 1 of \eq{odd1}.  The circles are the sieved  data  for $\bar pp$ scattering and the squares are the sieved data for $pp$ scattering for $\sqrt s\ge 6$ GeV. The solid curve ($\bar pp$)  and the dotted curve ($pp$) are $\chi^2$ cross section fits, corresponding to a simultaneous fit to cross sections and $\rho$-values   (Table \ref{table:ppoddfit}, of odderon 1) of  \eq{sigmapm1} and \eq{rhopm1}. 
}
\label{fig:sigmaodd1}
\end{figure}

\begin{figure}[h,t,b] 
\begin{center}
\mbox{\epsfig{file=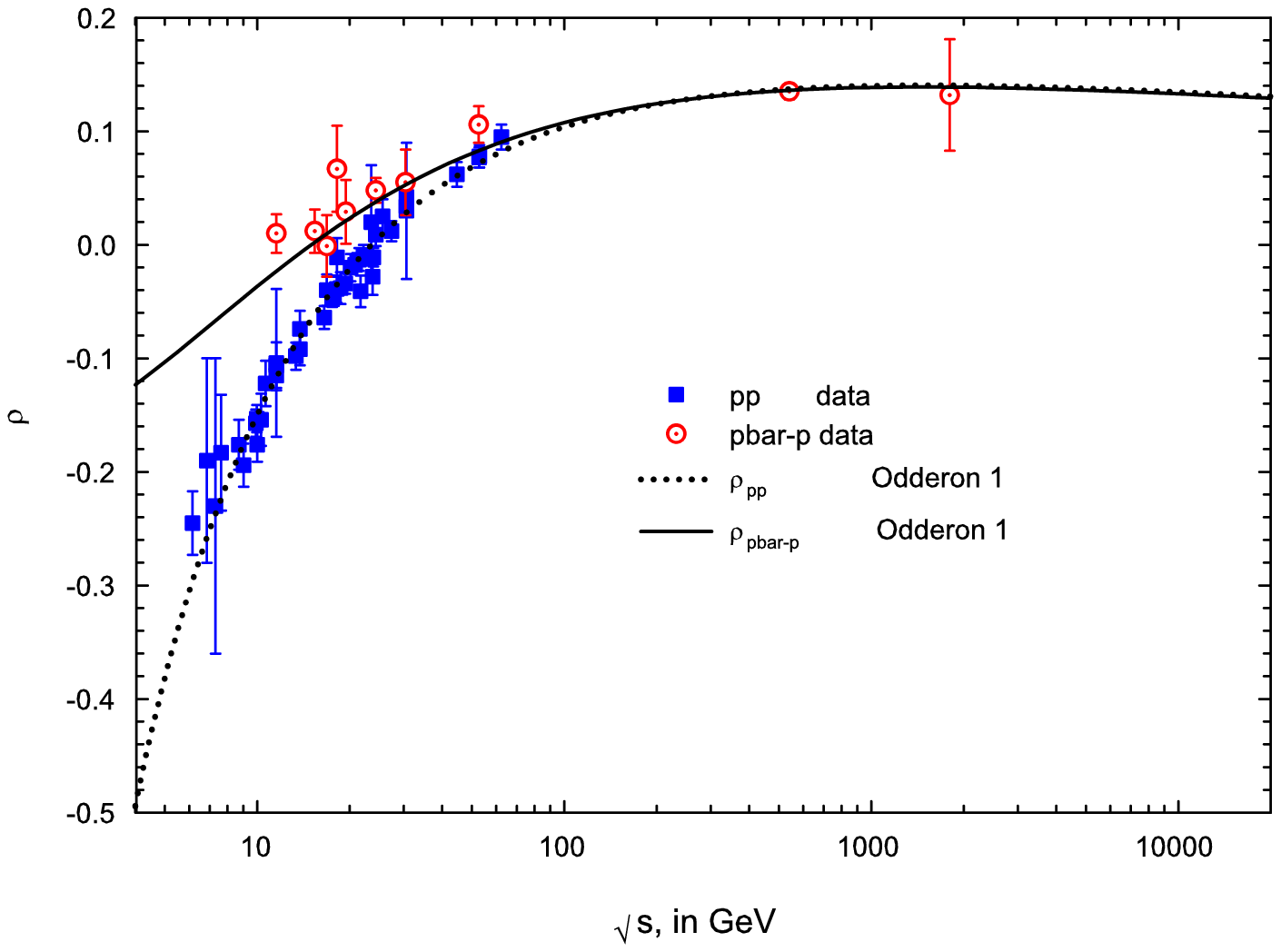,width=4.8in%
,bbllx=0pt,bblly=0pt,bburx=430pt,bbury=340pt,clip=%
}}
\end{center}
\caption[]{ \footnotesize
Odderon 1: The fitted $\rho$-values, $\rho_{\bar pp}$ and $\rho_{pp}$, vs. $\sqrt s$, in GeV, using the 4 constraints of Equations (\ref{deriveven}), (\ref{intercepteven}), (\ref{derivodd1}) and (\ref{interceptodd1}), for odderon 1 of \eq{odd1}.  The circles are the sieved  data  for $\bar pp$ scattering and the squares are the sieved data for $pp$ scattering for $\sqrt s\ge 6$ GeV. The solid curve ($\bar pp$)  and the dotted curve ($pp$) are $\chi^2$ cross section fits, corresponding to a simultaneous fit to cross sections and $\rho$-values   (Table \ref{table:ppoddfit}, of odderon 1) of \eq{sigmapm1} and \eq{rhopm1}. 
}
\label{fig:rhoodd1}

\end{figure}

\begin{figure}[h,t,b] 
\begin{center}
\mbox{\epsfig{file=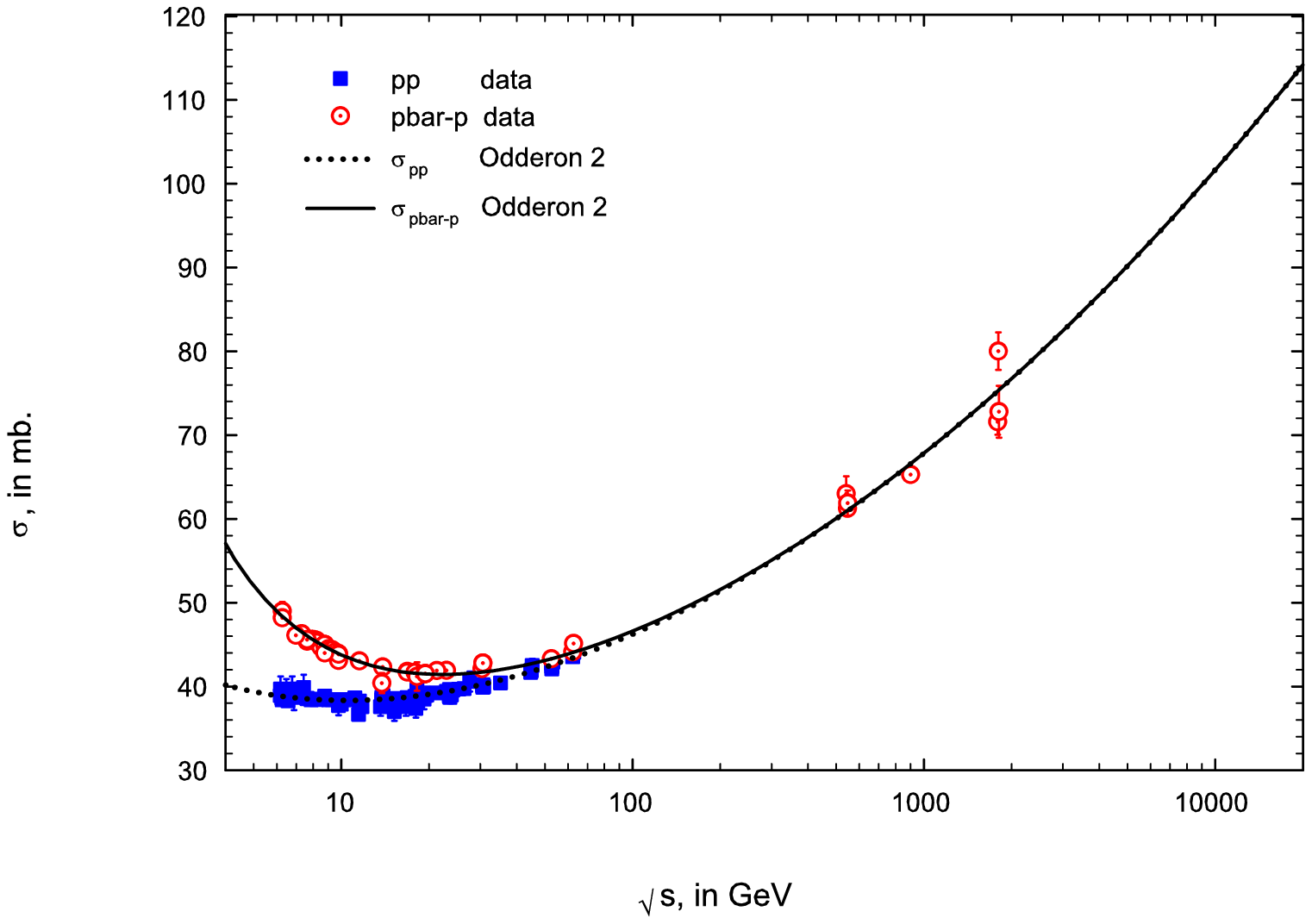,width=4.8in%
,bbllx=0pt,bblly=0pt,bburx=440pt,bbury=325pt,clip=%
}}
\end{center}
\caption[]{ \footnotesize
Odderon 2: The fitted total cross sections $\sigma_{\bar pp}$ and $\sigma_{pp}$ in mb, vs. $\sqrt s$, in GeV, using the 4 constraints of Equations (\ref{deriveven}), (\ref{intercepteven}), (\ref{derivodd2}) and (\ref{interceptodd2}), for odderon 2 of \eq{odd2}.  The circles are the sieved  data  for $\bar pp$ scattering and the squares are the sieved data for $pp$ scattering for $\sqrt s\ge 6$ GeV. The solid curve ($\bar pp$)  and the dotted curve ($pp$) are $\chi^2$ cross section fits, corresponding to a simultaneous fit to cross sections and $\rho$-values   (Table \ref{table:ppoddfit}, of odderon 2) of \eq{sigmapm2} and \eq{rhopm2}. 
}
\label{fig:sigmaodd2}
\end{figure}

\begin{figure}[h,t,b] 
\begin{center}
\mbox{\epsfig{file=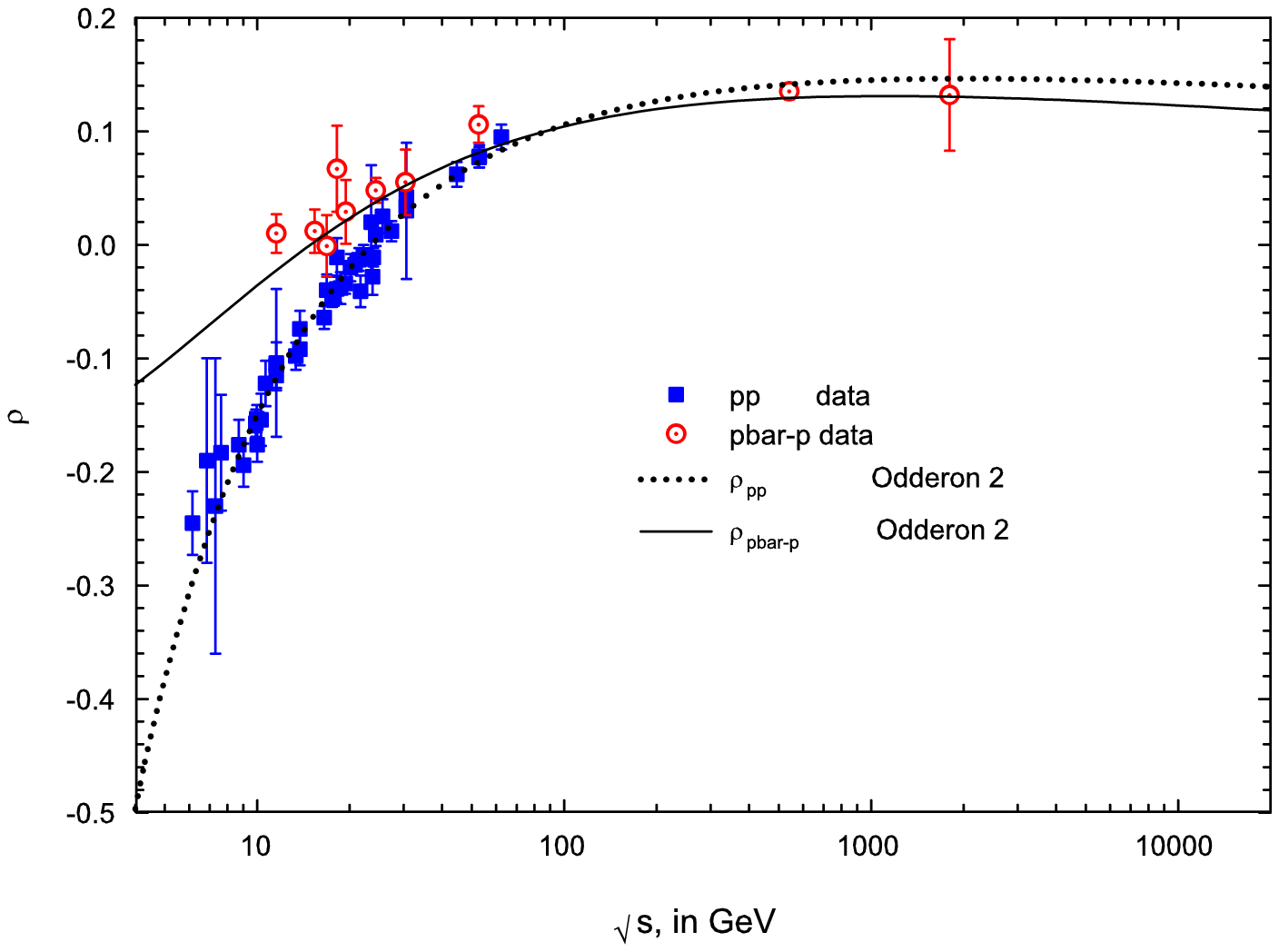,width=4.9in%
,bbllx=0pt,bblly=0pt,bburx=445pt,bbury=325pt,clip=%
}}
\end{center}
\caption[]{ \footnotesize
Odderon 2: The fitted $\rho$-values, $\rho_{\bar pp}$ and $\rho_{pp}$, vs. $\sqrt s$, in GeV, using the 4 constraints of Equations (\ref{deriveven}), (\ref{intercepteven}), (\ref{derivodd2}) and (\ref{interceptodd2}), for odderon 2 of \eq{odd2}.  The circles are the sieved  data  for $\bar pp$ scattering and the squares are the sieved data for $pp$ scattering for $\sqrt s\ge 6$ GeV. The solid curve ($\bar pp$)  and the dotted curve ($pp$) are $\chi^2$ cross section fits, corresponding to a simultaneous fit to cross sections and $\rho$-values   (Table \ref{table:ppoddfit}, of odderon 2) of  \eq{sigmapm2} and \eq{rhopm2}. 
}
\label{fig:rhoodd2}
\end{figure}

We summarize our results below:
\begin{itemize}
\item Odderon 0: Figure \ref{fig:sigmaodd0} shows the individual fitted cross sections (in mb) for $ pp$ and $\bar pp$ for  odderon 0 in Table \ref{table:ppoddfit}, plotted against the c.m. (center-of-mass) energy $\sqrt s$, in GeV. The data shown are the sieved data which have energies  $\sqrt s \ge 6$ GeV. The  fits to the data sample with $\delchimax=6$, corresponding to the dotted curve for $\bar pp$ and the solid curve for $pp$,  are excellent, yielding a total renormalized $\chi^2=201.2$, for 183 degrees of freedom, corresponding to a fit probability of $\sim0.2$. Figure \ref{fig:rhoodd0} shows the simultaneously  fitted $\rho$-values for $pp$ and $\bar pp$  for odderon 0 from Table \ref{table:ppoddfit}, plotted against the c.m. energy $\sqrt s$,  in GeV. The data shown are the sieved data with $\sqrt s \ge 6$ GeV. The solid curve for $\bar pp$ and the dotted curve for $pp$ fit the data reasonably well. It should be noted from Table \ref{table:ppoddfit} that the magnitude of odderon 0 is $\epsilon^{(0)}=-0.034\pm0.073$ mb, a very small coefficient. Indeed, it is  compatible with zero.
\item Odderon 1: Figure \ref{fig:sigmaodd1} shows the individual fitted cross sections (in mb) for $ pp$ and $\bar pp$ for  odderon 1 in Table \ref{table:ppoddfit}, plotted against the c.m. energy $\sqrt s$, in GeV. The data shown are the sieved data which have energies $\sqrt s \ge 6$ GeV. The  fits to the data sample with $\delchimax=6$, corresponding to the dotted curve for $\bar pp$ and the solid curve for $pp$,  are excellent, yielding a total renormalized $\chi^2=200.9$, for 183 degrees of freedom, corresponding to a fit probability of $\sim0.2$. Figure \ref{fig:rhoodd1} shows the simultaneously  fitted $\rho$-values for $pp$ and $\bar pp$  for odderon 1 from Table \ref{table:ppoddfit}, plotted against the c.m. energy $\sqrt s$,  in GeV. The data shown are the sieved data with $\sqrt s \ge 6$ GeV. The solid curve for $\bar pp$ and the dotted curve for $pp$ fit the data reasonably well. It should be noted from Table \ref{table:ppoddfit} that the magnitude of odderon 1 is $\epsilon^{(1)}=-0.0051\pm0.0077$ mb, a very tiny coefficient which is again compatible with zero.
\item Odderon 2: Figure \ref{fig:sigmaodd2} shows the individual fitted cross sections (in mb) for $ pp$ and $\bar pp$ for  odderon 2 in Table \ref{table:ppoddfit}, plotted against the c.m. energy $\sqrt s$, in GeV. The data shown are the sieved data which have energies $\sqrt s \ge 6$ GeV. The  fits to the data sample with $\delchimax=6$, corresponding to the dotted curve for $\bar pp$ and the solid curve for $pp$,  are excellent, yielding a total renormalized $\chi^2=196.1$, for 183 degrees of freedom, corresponding to a fit probability of $\sim0.2$. Figure \ref{fig:rhoodd2} shows the simultaneously  fitted $\rho$-values for $pp$ and $\bar pp$  for odderon 2 from Table \ref{table:ppoddfit}, plotted against the c.m. energy $\sqrt s$,  in GeV. The data shown are the sieved data with $\sqrt s \ge 6$ GeV. The solid curve for $\bar pp$ and the dotted curve for $pp$ fit the data reasonably well.  It should be noted from Table \ref{table:ppoddfit} that the magnitude of odderon 2 is $\epsilon^{(2)}=0.0042\pm0.0019$ mb, a very tiny coefficient which is only about two standard deviations from zero.
\end{itemize}

In Table \ref{table:predictions}, we make predictions of  total cross sections and $\rho$-values for $\bar pp$ and $pp$ scattering for odderon 2 of Table \ref{table:ppoddfit}. Only for  {\em very} high energies  above $\sqrt s=14$ TeV  is there any appreciable difference between $\rho_{\bar pp}$ and $\rho_{pp}$, as seen in Fig. \ref{fig:rhoodd2}. In fact, the results of all 3 fits are very close to what was found in ref. \cite{bhpp}, where there were no odderon amplitudes, but had virtually identical $\chi^2$/d.f. 

These new upper limits on odderon amplitudes are to be contrasted to the analysis made in 1985 by Block and Cahn\cite{bc}, where they found $\epsilon^{(0)}=-0.25\pm0.13$ mb, $\epsilon^{(1)}=-0.11\pm0.04$ mb and $\epsilon^{(2)}=-0.04\pm0.02$ mb, which were about two standard deviations from zero, but with errors of almost 2 to 10 times larger than the limits found in this note.  Our marked increase in present accuracy is attributable to the use of the 4 analyticity constraints\cite{analyticity} employed in the present analysis, as well as to the use of the improved sieved data set\cite{bhpp,sieve}, which also has  higher energy points than were available in 1985.

In conclusion, the magnitude of all three odderon amplitudes, $\epsilon^{(0)}=-0.034\pm0.073$ mb, $\epsilon^{(1)}=-0.00051\pm0.0077$ mb and $\epsilon^{(2)}=0.0042\pm0.0019$ mb, in comparison to all of the other amplitudes found in the fit---typically of the order of 1.5 to 40 mb---are very tiny. Indeed, all 3 are compatible with zero and we now can set new upper limits a factor of 2 better for $\epsilon^{(0)}$, a factor of 5 better for $\epsilon^{(1)}$ and a factor of 10 better for the maximum odderon $\epsilon^{(2)}$. An accurate measurement of the $\rho$-value at the LHC, where Block and Halzen\cite{bhpp} predict $\rho_{pp}=0.132\pm0.001$ when odderon amplitudes are zero and our prediction from  from Table \ref{table:predictions} is $\rho_{pp}= 0.141\pm0.005$, would really constrain the maximal odderon amplitude $\epsilon^{(2)}$. 

\section*{Acknowledgments}One of us (MMB) would like to acknowledge the hospitality of the Aspen Center for Physics, Aspen, CO,  during the preparation of this work. One of us (KK) was supported in part by the  Department of Energy contract DE-FG02-91Er40688 Task A.

\newpage

\begin{table}[h,t]                   
%
\def\arraystretch{1.5}            
%
\begin{center}
\begin{tabular}[b]{|l|c|c}
      \hline
     $\nu_0$, lab transition energy \ \ \ \ \ \ \ \ \ \  (GeV)&7.59  					\\ 		 
	$\rightarrow \ \sqrt s_0$, c.m. transition energy \ \ (GeV)&4\\
		\hline
	$\sigma_+(\nu_0)$ \ \ \ \ \ \ \ \ \ \  \    (mb)&40.18\\
      $\sigma_-(\nu_0)$  \ \ \ \ \ \ \ \ \ \  \   (mb)&56.99\\

      $\left(\frac{d\sigma_+}{d(\nu/m)}\right)_{\nu=\nu_0}$\ \ \ \ (mb)
		&-0.2305 	\\ 
      $\left(\frac{d\sigma_-}{d(\nu/m)}\right)_{\nu=\nu_0}$\ \ \ \ (mb)
	    &-1.446\\
      \hline\hline
\multicolumn{1}{|c|}{Minimum fitting energy} &\multicolumn{1}{c|}{}\\
\hline
$\nu_{\rm min}$,\ \ \  \ \ \ \ \ lab minimum energy \ (GeV)&18.25\\
$\rightarrow \ \sqrt {s_{\rm min}}$, c.m. minimum energy\   (GeV)&6.0\\
     \hline
\multicolumn{2}{c}{$m$ is the proton mass and $\nu$ is the laboratory proton energy}
\end{tabular}
\end{center}
     \caption{\protect\small The transition energy parameters and minimum fitting energy used for constraining $pp$ and  $\bar pp $ scattering. Taken from ref. \cite{bhpp}.\label{table:transitionparameters}
}
\end{table}

\begin{table}[h,t]                   
%
\def\arraystretch{1.5}            
\begin{center}				  
\begin{tabular}[b]{|l||c|c|c|}
\hline
Parameters&odderon 0&odderon 1&odderon 2\\
      \hline
	{}&\multicolumn{3}{|c|}{ Even Amplitude}\\
	\cline{1-4}
      $c_0$\ \ \   (mb)&$37.38$ &$37.24$&37.09\\ 
      $c_1$\ \ \   (mb)&$-1.460\pm0.065$ &$-1.415\pm0.073$&$-1.370\pm 0.0074$\\ 
	$c_2$\ \ \ \   (mb)&$0.2833\pm0.0060$&$0.2798\pm0.0064$&$0.2771\pm0.0064$\\
      $\beta_{\cal P'}$\ \   (mb)&$37.02$ &$37.20$&37.39\\ 
      $\mu$&$0.5$ &$0.5$&0.5\\ 
	$f_+(0)$ (mb GeV)&$-0.075\pm0.75$&$-0.050\pm 0.59$&$-.073\pm 0.58$\\
      \hline
	{}&\multicolumn{3}{|c|}{ Odd Amplitude}\\
	\hline
      $\delta$\ \ \   (mb)&$-28.56$ &$-28.53$&-28.49\\
      $\alpha$&$0.415$ &$0.416$&0.416\\ 
$\epsilon^{(j)}$ (mb)$,\quad j=0,1,2$&$-0.034\pm0.073$&$-0.0051\pm0.0077$&$0.0042\pm0.0019$\\
     	\hline
	\hline
	$\chi^2_{\rm min}$&181.3&181.1&176.7\\
	${\cal R}\times\chi^2_{\rm min}$&201.2&200.9&196.1\\ 
	degrees of freedom (d.f.)&183&183&183\\
\hline
	${\cal R}\times\chi^2_{\rm min}$/d.f.&1.099&1.098&1.071\\
\hline
\end{tabular}
\end{center}
     \caption{\protect\small The fitted results for a 4-parameter $\chi^2$ fit using odderons 0, 1 and 2, with $\sigma\sim\ln^2 s$, to the total cross sections and $\rho$-values for $pp$ and $\bar pp$ scattering. The renormalized $\chi^2_{\rm min}$ per degree of freedom,  taking into account the effects of the $\delchimax=6$ cut, is given in the row  labeled ${\cal R}\times\chi^2_{\rm min}$/d.f. The errors in the fitted parameters have been multiplied by the appropriate $r_{\chi2}$. For details on the renormalization of the errors by $r_{\chi2}$ and the renormalization of $\chi^2_{\rm min}$ by ${\cal R}$, see ref. \cite{sieve}. 
\label{table:ppoddfit}}
\end{table}
\def\arraystretch{1}  
\begin{table}[h,t]                   
%
\def\arraystretch{1.5}            
\begin{center}
\begin{tabular}[b]{|l||c|c||c|c||}
    \cline{1-5}
      \multicolumn{1}{|l||}{ $\sqrt s$, in GeV}
      &\multicolumn{1}{c|}{$\sigma_{\bar pp}$, in mb}
      &\multicolumn{1}{c||}{$\rho_{\bar p p}$}&\multicolumn{1}{c|}{$\sigma_{ pp}$, in mb}&\multicolumn{1}{c||}{$\rho_{pp}$}\\

      \hline\hline
	300&$55.14\pm0.20$&$0.125\pm0.003$&$54.82\pm0.20$&$0.134\pm 0.003$\\
\hline	
	540&$60.89\pm0.29$&$0.129\pm0.004$&$60.59\pm0.29$&$0.141\pm0.003$\\\hline
 	1,800&$75.19\pm0.50$&$0.130\pm0.001$&$74.87\pm0.52$&$0.146\pm0.004$\\\hline    
 	14,000&$107.1\pm1.1$&$0.121\pm0.005$&$106.6\pm1.1$&$0.141\pm0.005$\\\hline    
 	50,000&$131.55\pm1.5$&$0.112\pm0.006$&$131.1\pm1.6$&$0.134\pm0.005$\\\hline
 	100,000&$146.39\pm1.8$&$0.108\pm0.006$&$145.9\pm1.9$&$0.131\pm0.005$\\\hline
\end{tabular}
\end{center}
     \caption{\protect\small Predictions of high energy $\bar pp$ and $pp$ total  cross sections and $\rho$-values for odderon 2,  from Table \ref{table:ppoddfit}.\label{table:predictions}
}
\end{table}

\def\arraystretch{1}  
\def\arraystretch{1}  

%
%
%

\newpage
\begin{thebibliography}{99} 
\frenchspacing
\bibitem{nicolescu1}
L.~Lukaszuk and B.~Nicolescu, Lett. Nuovo Cimento {\bf 8}, 405 (1973).
\bibitem{nicolescu2}
K. Kang and B.~Nicolescu, Phys. Rev. D {\bf 11},2461 (1975). This was the first group to apply derivative dispersion relations to the derivation of real analytic amplitudes.
\bibitem{joynson}
D. Joynson et al, Nuovo Cimento A{\bf 30}, 345 (1975).
\bibitem{bk} M. M. Block and K. Kang, Int. J. Mod. Phys. A{\bf 20}, 2781 (2005).
\bibitem{cudell}
J. R. Cudell {\em et al.}, Phys. Rev. D {\bf 65}, 074024 (2002); (COMPETE Collaboration) Phys. Rev. Lett. {\bf 89}, 201801 (2002).
\bibitem{igiandishidapip} 
K. Igi and M. Ishida, Phys. Rev. D {\bf{ 66}}, 034023 (2002).
\bibitem{igiandishidapp} 
K. Igi and M. Ishida, Phys. Lett.  {\bf B622}, 286 (2005).
\bibitem{blackhole}  
K. Kang and H. Nastase, Phys. Rev. D {\bf 72}, 106003 (2005); Phys. Lett. B {\bf 624}, 125 (2005).
\bibitem{BH}M. M. Block and F. Halzen,  {\bf hep-ph0405174} (2004); Phys. Rev. D {\bf 70}, 091901 (2004).
\bibitem{bhpp} 
M. M. Block and F. Halzen, Phys. Rev D {\bf 72}, 036006 (2005).
\bibitem{bc} 
 M.~M.~Block and R.~N.~Cahn, Rev.~Mod.~Phys.~{\bf 57}, 563 (1985).
\bibitem{compton} M. M.  Block, Phys. Rev. D {\bf 65}, 116005 (2002).
\bibitem{analyticity} M. M. Block, {\bf hep-ph0601210} (2006).
\bibitem{gilman} 
For the reaction $\gamma +p\rightarrow\gamma + p$, it is fixed as the Thompson scattering limit $f_+(0)=-\alpha/m=-3.03\  \mu {\rm b\  GeV}$ [see
M. Damashek and F. J. Gilman, Phys. Rev. D {\bf 1}, 1319 (1970)].
\bibitem{sieve}M. M. Block, {\bf physics/0506010} (2005);  Nucl. Inst. and Meth. A. {\bf 556}, 308 (2006).
\bibitem{pdg} 
Particle Data Group, K. Hagiwara  et al., Phys. Rev. D {\bf 66}, 010001 (2002).
%
\end{thebibliography}
\end{document}